\documentclass[aps,showpacs,preprint,footinbib,preprintnumbers]{revtex4}
\usepackage{amssymb}
\usepackage{amsmath}
\usepackage{amsfonts}
\usepackage{bm}
\usepackage{here}
\usepackage{graphicx}

\newcommand{\be}{\begin{equation}}
\newcommand{\ee}{\end{equation}}
\newcommand{\ba}{\begin{eqnarray}}
\newcommand{\ea}{\end{eqnarray}}
\newcommand{\nn}{\nonumber}

\newcommand{\pr}{\prime}

\begin{document}

\title{The $\Lambda(1405)$ state in a chiral unitary approach with off-shell corrections to dimensional regularized loop functions}

\author{ Fang-Yong Dong$^{1}$, Bao-Xi Sun$^{1,2}$ and Jing-Long Pang$^{2}$}

\affiliation{${}^{1}$College of Applied Sciences, Beijing University
of Technology, Beijing 100124, China}

\affiliation{${}^{2}$Department of Physics, Peking University,
Beijing 100871, China}

\begin{abstract}
The Bethe-Salpeter equation is solved in the framework of unitary
coupled-channel approximation by using the pseudoscalar meson-baryon
octet interaction. The loop function of the intermediate meson and
baryon is deduced in a dimensional regularization
scheme, where the relativistic kinetic effect and off-shell corrections are taken into account.
According to the experimental data at the $K^- p$ threshold, the subtraction constants in the loop function are determined.
The squared amplitude is suppressed strongly and only one $\Lambda(1405)$ state is generated dynamically in the strangeness
$S=-1$ and isospin $I=0$ sector.

\end{abstract}


\maketitle

\section{Introduction}

There are some different views about the structure of
$\Lambda(1405)$, and its structure has been challenging the standard
view of baryons made of three quarks for decades. Some theorists
think that the $\Lambda(1405)$ could be a kind of molecular state
arising from the interaction of the $\pi\Sigma$ and $\bar{K}N$
channels~\cite{Dalitz:1960,3Dalitz:1967,Kisslinger:2009dr}.
Furthermore, the number of poles on the complex energy plane is also a puzzle for the
$\Lambda(1405)$ resonance. Some people predict one pole corresponds
to the $\Lambda(1405)$
resonance~\cite{4Zychor:2008,5Isgur:1978,Dalitz:1960}, while the
calculation results in the unitary coupled-channel approximation
show there are two poles in the $1400$MeV
region\cite{17Oller:2001,8Jido:2003,10Geng:2007,15Oset:2002,16Hyodo:2003}.
Two $\Lambda(1405)$ states are claimed for the first time in Ref.~\cite{17Oller:2001},
which are explained as combinations of a single state
and a octet state when the $SU(3)$ symmetry breaks up
\cite{8Jido:2003}.
It is reported that only one resonance state is observed around
$1400$MeV~\cite{11Ciepl:2010,12Maryam:2013,13Dalitz:1991,14Gideon}.
However, in a modern $K^+ \Sigma \pi$ photoproduction experiment made by CLAS
Collaboration, some signature
effect for a two-pole picture of the isospin $I=0$ $\Lambda(1405)$ has been found\cite{Moriya:2013eb}.
%


In the past years, the energy shift and width of the 1s state of the Kaonic hydrogen are measured precisely
in the SIDDHARTA experiment at $DA \Phi NE$\cite{SIDDHARTA}, which provide a constraint on the determination of parameters in the calculation of unitary coupled-channel approximation.
The new experimental data have stimulated the theoretical study
on the Kaon-nucleon interaction and the properties of
the $\Lambda(1405)$ particle\cite{Ikeda:2012au, Mai:2012dt, Guo:2012vv, Mai:2014xna}, where the Bethe-Salpeter equation
is solved in the unitary coupled-channel approximation, and the two-pole picture of the $\Lambda(1405)$
particle is stressed.
The direct comparison of the most recent approaches is made in Ref.~\cite{Cieply:2016jby}.
However, a loop function of the meson and the
baryon in the on-shell factorization is used when the Bethe-Salpeter
equation is solved, and some important elements might be eliminated
in this approximation, which would result in the uncertainty of the
calculation.
In this article, we calculate the loop function in a dimensional regularization scheme, and then solve the
Bethe-Salpeter equation in the unitary coupled-channel
approximation.

This manuscript is organized as follows. In
Section~\ref{sect:frame}, the framework of unitary coupled-channel
approximation is discussed, especially, the
loop function of the pseudoscalar meson and the baryon octet is
obtained in the dimensional regularization.
In Section~\ref{sect:fits},
the parameters in the unitary coupled-channel approximation are determined according to the experimental data at the
$K^- p$ threshold.
In
Sections~\ref{sect:I0} and~\ref{sect:I1},
The cases with isospin $I=0$ and $I=1$ are calculated, respectively.
Some discussions and a conclusion are given in
Section~\ref{sect:summary}.

\section{Framework}
\label{sect:frame}

The lowest order meson-baryon chiral Lagrangian is given as
\cite{Pich1995,Ecker1995,Meissner1995}
\begin{align}
   L={}&\langle \bar{B}i\gamma^{\mu}\frac{1}{4f^{2}}
   [(\Phi\partial_{\mu}\Phi-\partial_{\mu}\Phi\Phi)B
   -B(\Phi\partial_{\mu}\Phi-\partial_{\mu}\Phi\Phi)]\rangle ,
   \label{eq:Lagrangian}
\end{align}
where the symbol $\langle...\rangle$ denotes the trace of matrices
in the $SU(3)$ space. The matrices of the pseudoscalar meson and the
baryon octet are given as follows

\begin{align}
\Phi={}&
\begin{pmatrix}
\frac{1}{\sqrt{2}}\pi^{0}+\frac{1}{\sqrt{6}}\eta & \pi^{+} & K^{+}\\
\pi^{-} &  -\frac{1}{\sqrt{2}}\pi^{0}+\frac{1}{\sqrt{6}}\eta & K^{0}\\
K^{-} & \bar{K}^{0} & -\frac{2}{\sqrt{6}}\eta
\end{pmatrix}
 \label{eq:mesons matrices}
\end{align}
and
\begin{align}
B={}&
\begin{pmatrix}
\frac{1}{\sqrt{2}}\Sigma^{0}+\frac{1}{\sqrt{6}}\Lambda & \Sigma^{+} & p\\
\Sigma^{-} &  -\frac{1}{\sqrt{2}}\Sigma^{0}+\frac{1}{\sqrt{6}}\Lambda & n\\
\Xi^{-} & \Xi^{0} & -\frac{2}{\sqrt{6}}\Lambda
\end{pmatrix}.
 \label{eq:baryons matrices}
\end{align}

Ten coupled channels are considered in the case of pseudoscalar
meson-baryon octet scattering processes at the low energy region,
which are
$K^{-}p$,$\bar{K}^{0}n$,$\pi^{0}\Lambda$,$\pi^{0}\Sigma^{0}$,$\pi^{+}\Sigma^{-}$,$\pi^{-}\Sigma^{+}$,\\$\eta\Lambda$,$\eta\Sigma^{0}$,$K^{+}\Xi^{-}$and$K^{0}\Xi^{0}$~\cite{8Jido:2003}.

The potential of the baryon octet and the pseudoscalar meson can be
obtained from the lowest order meson-baryon chiral Lagrangian in
Eq.~(\ref{eq:Lagrangian}), and can be written as
\be
\label{eq:v170123}
V_{ij}=-C_{ij}\frac{1}{4f^{2}}\bar{U}(p_2)\gamma_\mu U(p_1)(k_1^\mu+k_2^\mu),
\ee
where $p_1,~p_2$ $(k_1,~k_2)$ are the initial and final momenta of baryons $(mesons)$, respectively.
In the case of low energies, the three-momenta of baryons and mesons can be neglected, and thus only the $\gamma^0$ component is relevant, i.e.,
\be
\bar{U}(p_2)\gamma_\mu U(p_1)\approx \left(\frac{M_{i}+E}{2M_{i}}\right)^{\frac{1}{2}}
\left(\frac{M_{j}+E^{\prime}}{2M_{j}}\right)^{\frac{1}{2}},
\ee
where $M_{i}$ and $M_{j}$ denote the initial
and final baryon masses, respectively, while $E$ and $E^\prime$ stand for the
initial and final baryon energies in the center of mass frame,
respectively. The spin orientation of the initial baryon is the same as that of the final baryon. Moreover,
\be
k_1^0+k_2^0\approx 2\sqrt{s}-M_{i}-M_{j},
\ee
with $\sqrt{s}$ the total energy of the system in the center of mass frame.

Therefore, if and only if the external particles are all on-shell,
the potential in Eq.~(\ref{eq:v170123}) takes a simple form as follow~\cite{8Jido:2003}:
\begin{align}
V_{ij}={}&-C_{ij}\frac{1}{4f^{2}}(2\sqrt{s}-M_{i}-M_{j})
\left(\frac{M_{i}+E}{2M_{i}}\right)^{\frac{1}{2}}
\left(\frac{M_{j}+E^{\prime}}{2M_{j}}\right)^{\frac{1}{2}},
\label{eq:V_simple}
\end{align}
where the coefficient $C_{ij}$ is shown in Table~\ref{TABLE1}, and
the decay constant $f=1.123f_{\pi}$~\cite{15Oset:2002} with the pion
decay constant $f_{\pi}=92.4$MeV.

The scattering amplitude can be constructed by solving the Bethe-Salpeter equation
\ba
\label{eq:1701231254}
T(p_1,k_2;p_2,k_2)&=&V(p_1,k_2;p_2,k_2) \nn \\
&+&i\int \frac{d^4 q}{(2\pi)^4} V(p_1,k_1;q,p_1+k_1-q) S(q) \Delta(p_1+k_1-q)T(q,p_1+k_1-q;p_2,k_2) \nn \\
&=&V(p_1,k_2;p_2,k_2) \nn \\
&+&i\int \frac{d^4 q}{(2\pi)^4} V(p_1,k_1;q,p_1+k_1-q) S(q) \Delta(p_1+k_1-q)V(q,p_1+k_1-q;p_2,k_2) \nn \\
&+&...,
\ea
where
%
%
the propagators of the intermediate baryon and meson can be written as
 $iS(q)=i/(\rlap{/}q+M_{l}+i\varepsilon)$ and $i\Delta(p_1+k_1-q)=i/[(p_1+k_1-q)^{2}-m_{l}^{2}+i\epsilon]$,
 respectively\cite{Bruns:2010sv}.
If the potential $V(p_1,k_1;q,p_1+k_1-q)$ in Eq.~(\ref{eq:1701231254}) is divided into a on-shell part and an off-shell part, the off-shell part of the potential $V(p_1,k_1;q,p_1+k_1-q)$ would be proportional to the on-shall
part, and can be absorbed into the on-shell part of the potential if a suitable renormalization of coupling constants is preformed. Therefore, only the on-shell part of the potential of the baryon and the meson is necessary to be taken into account when the Bethe-Salpeter equation is solved. Thus the Bethe-Salpeter equation in Eq.~(\ref{eq:1701231254}) is simplified as
\be
T=V+VGT,
\ee
or
\begin{align}
T={}&[1-VG]^{-1}V, \label{eq:amplitude}
\end{align}
where the loop
function of a baryon and a meson $G$ is a diagonal matrix, 
i.e., $G_{ln}=G_l \delta_{ln}$, and the diagonal element $G_l$
can be written
as
\begin{align}
G_{l}={}&i\int\frac{d^{4}q}{(2\pi)^{4}}\frac{\rlap{/}q+M_{l}}{q^{2}-M_{l}^{2}+i\epsilon}\frac{1}{(P-q)^{2}-m_{l}^{2}+i\epsilon}
 \label{eq:G}
\end{align}
with $P=p_1+k_1$ the total momentum of the system, $m_{l}$ the meson mass,
and $M_{l}$ the baryon mass, respectively.
More detailed discussion on how to transform the Bethe-Salpeter equation from an integral form to an algebra form  can be found in Refs.~\cite{7Oset:1998,Bruns:2010sv}.

In Ref.~\cite{7Oset:1998}, the loop function $G$ in Eq.~(\ref{eq:G}) is calculated numerically by setting the maximum three-momentum $q_{max}$, which is called the momenum cut-off method. However,
in Ref.~\cite{17Oller:2001}, a dimensional regularization form of
the loop function $G$ is deduced with the on-shell approximation
\be
\rlap{/}q+M_{l}=2M_{l},
\ee
which is valid only when it is applied on the Dirac spinor $U(q)$.

In the on-shell factorization
approximation, the loop function is denoted as
\begin{eqnarray}
G^\prime_{l}(s) &=&
i\int\frac{d^{d}q}{(2\pi)^{4}}\frac{2M_{l}}{q^{2}-M_{l}^{2}+i\epsilon}
\frac{1}{(P-q)^{2}-m_{l}^{2}+i\epsilon} \nn \\
&=& \frac{2 M_l}{16 \pi^2} \left\{ a_l + \ln
\frac{M_l^2}{\mu^2} + \frac{m_l^2-M_l^2 + s}{2s} \ln
\frac{m_l^2}{M_l^2} + \right. \nonumber \\ & &  \phantom{\frac{2
M}{16 \pi^2}} + \frac{\bar{q}_l}{\sqrt{s}} \left[
\ln(s-(M_l^2-m_l^2)+2\bar{q}_l\sqrt{s})+
\ln(s+(M_l^2-m_l^2)+2\bar{q}_l\sqrt{s}) \right. \nonumber  \\
& & \left. \phantom{\frac{2 M}{16 \pi^2} +
\frac{\bar{q}_l}{\sqrt{s}}} \left. \hspace*{-0.3cm}-
\ln(-s+(M_l^2-m_l^2)+2\bar{q}_l\sqrt{s})-
\ln(-s-(M_l^2-m_l^2)+2\bar{q}_l\sqrt{s}) \right] \right\} \
\label{eq:gpropdr}
\end{eqnarray}
with $\mu=630$MeV the scale of dimensional regularization and the
symbol $a_{l}$ the subtraction constant.

In Eq.~(\ref{eq:gpropdr}), $\bar{q}_l$ denotes the three-momentum of
the meson or the baryon in the center of mass frame
and is given by
\begin{equation}
\bar{q}_l=\frac{\lambda^{1/2}(s,m_l^2,M_l^2)}{2\sqrt{s}}
=\frac{\sqrt{s-(M_l+m_l)^2}\sqrt{s-(M_l-m_l)^2}}{2\sqrt{s}},
\end{equation}
with $\lambda$ the triangular function.

Actually, the loop function in Eq.~(\ref{eq:G}) can be calculated
 in the dimensional regularization without the on-shell
approximation taken into account. Thus the loop function takes the
form of
\begin{equation}
\begin{aligned}
G_{l}={}&\frac{\gamma_{\mu}
P^{\mu}}{32P^{2}\pi^{2}}\left[(a_{l}+1)(m_{l}^{2}-M_{l}^{2})+(m_{l}^{2}\ln\frac{m_{l}^{2}}{\mu^{2}}-M_{l}^{2}\ln\frac{M_{l}^{2}}{\mu^{2}})\right]\\&
+\left(\frac{\gamma_{\mu}
P^{\mu}[P^{2}+M_{l}^{2}-m_{l}^{2}]}{4P^{2}M_{l}}+\frac{1}{2}\right)G_{l}^{\prime}.
\end{aligned} \label{eq:Our G}
\end{equation}
Since the total three-momentum $\vec{P}=0$ in the center of mass
frame, only the $\gamma_{0} P^{0}$ parts remain in Eq.~(\ref{eq:Our
G}).
The external particles in the potential of the baryon and the meson in Eq.~(\ref{eq:V_simple}) are all on-shell,
so the anti-particle is not included in the intermediate states
when the Bethe-Salpeter equation is solved.
Thus
$\gamma_{0}
P^{0}$ can be treated as the total energy of the system
$P^{0}=\sqrt{s}$ in Eq.~(\ref{eq:Our G}). Therefore, the loop
function in Eq.~(\ref{eq:Our G}) is simplified as
\begin{equation}
\begin{aligned}
G_{l}={}&\frac{\sqrt{s}}{32\pi^{2}s}\left[(a_l+1)(m_{l}^{2}-M_{l}^{2})+(m_{l}^{2}\ln\frac{m_{l}^{2}}{\mu^{2}}-M_{l}^{2}\ln\frac{M_{l}^{2}}{\mu^{2}})\right]\\&
+\left(\frac{s+M_{l}^{2}-m_{l}^{2}}{4M_{l}\sqrt{s}}+\frac{1}{2}\right)G_{l}^{\prime}
\end{aligned}\label{eq:Our G result}
\end{equation}
Apparently, some off-shell corrections have been included in the revised form of the loop function
in Eq.~(\ref{eq:Our G result}), which can be regarded as a kind of relativistic kinetic
effect of the loop function.



Assuming the amplitudes near the pole to behave as
\begin{equation} \label{eq:GIJ}
T_{ij}=\frac{g_{i}g_{j}}{z-z_{R}}
\end{equation}
with $z_{R}$ the position of the pole on the complex $\sqrt{s}$
plane, and $j$ and $i$ being the initial and final channels,
respectively, we can obtain the size of the coupling constants $g_i$ by
evaluating the residues of the diagonal elements $T_{ii}$. When the
strongest coupled channel is determined, the coupling constants to the other
channels, $g_j$, can be evaluated with the residues of the
non-diagonal elements $T_{ij}$ and the largest coupling constant $g_i$ by
using Eq.~(\ref{eq:GIJ}).


\section{Experimental data and parameter fits}
\label{sect:fits}

The energy shift and the width of the 1s state of the Kaonic hydrogen measured by the SIDDHARTA Collaboration are given as
\be
\Delta E=283\pm36\pm6 eV,
\ee
and
\be
\Gamma=541\pm89\pm22 eV,
\ee
respectively\cite{SIDDHARTA}.
These results would supply a constraint to the parameter fit when the Bethe-Salpeter equation is solved.

In order to fit the experimental data at the $K^- p$ threshold with the same formula as in Ref.~\cite{Ikeda:2012au},
the potential of the baryon octet and the
pseudoscalar meson in Eq.~(\ref{eq:V_simple}) has to be multiplied by a factor of $\sqrt{M_i M_j}$, where $M_i$
and $M_j$ denote the masses of the initial and final baryons, respectively, i.e.,
\be
\label{eq:1704031}
\tilde{V}_{ij}=V_{ij} \sqrt{M_i M_j}.
\ee
In the meantime, the loop function in Eq.~(\ref{eq:Our G result}) is divided by the mass of the intermediate
baryon $M_l$,
i.e.,
\be
\label{eq:1704032}
\tilde{G}_l=G_l/M_l.
\ee
Since the pole appears when the determinant $|1-VG|=0$, the modification in Eqs.~(\ref{eq:1704031})
and (\ref{eq:1704032}) will not affect the pole position on the complex $\sqrt{s}$ plane significantly.

The forward scattering amplitude $f_{ij}$ is related to the T-matrix,
\be
f_{ij}(\sqrt{s})=\frac{1}{8\pi\sqrt{s}}\tilde{T}_{ij}(\sqrt{s}),
%
\ee
with $\tilde{T}=[1-\tilde{V} \tilde{G}]^{-1} \tilde{V}$,
and the $K^- p$ scattering length can be defined by the $K^- p$ elastic scattering amplitude at threshold,
\be
a(K^{-}p)=f_{11}(\sqrt{s}=M_{K^{-}}+m_{p}),
\label{eq:scattering length}
\ee
which is a complex number when the inelastic channels
are taken into account.

The energy shift and width of the 1s state of the Kaonic hydrogen are related to the $K^- p$ scattering length,  which can be written as\cite{Meissner:2004jr}
\be
\Delta E-i\frac{1}{2}\Gamma=-2\alpha^{3}\mu_{\gamma}^{2}a(K^{-}p)\left[ 1+2\alpha\mu_{\gamma}(1-ln\alpha)a(K^{-}p)\right],
\label{eq:relationship E G A}
\ee
with $\alpha$ the fine structure constant and
$\mu_{\gamma}=\frac{m_{K^{-}} M_{p}}{m_{K^{-}}+M_{p}}$ the $K^- p$ reduced mass.

Moreover, the branching ratios at the $K^- p$ threshold defined as
\be
\gamma=\frac{\Gamma\left(K^{-}p\rightarrow \pi^{+}\Sigma^{-} \right)}{\Gamma\left( K^{-}p\rightarrow \pi^{-}\Sigma^{+} \right)}={}\frac{\sigma_{51}}{\sigma_{61}},
\ee

\be
\label{eq:Rn}
Rn=\frac{\Gamma\left(K^{-}p\rightarrow \pi^{0}\Lambda \right)}{\Gamma\left( K^{-}p\rightarrow \text{neutral states} \right)}=\frac{\sigma_{31}}{\sigma_{31}+\sigma_{41}},
\ee
and
\be
\label{eq:Rc}
Rc=\frac{\Gamma\left(K^{-}p\rightarrow \pi^{+}\Sigma^{-},\pi^{-}\Sigma^{+} \right)}{\Gamma\left( K^{-}p\rightarrow \text{all inelastic channels} \right)}=\frac{\sigma_{51}+\sigma_{61}}{\sigma_{51}+\sigma_{61}+\sigma_{31}+\sigma_{41}},
\ee
respectively. All partial cross sections $\sigma_{ij}$ are calculated at the $K^- p$ threshold.
\be
\sigma_{ij}=\frac{\bar{q}_{i}}{\bar{q}_{j}}\frac{\left\vert \tilde{T}_{ij}\right\vert^{2}}{16\pi s},
\ee
where $\bar{q}_j$ and $\bar{q}_i$ are the three-momentum of the initial and final states in the center of mass frame, respectively.

The values of three branching ratios are taken from Ref.~\cite{Tovee:1971ga,Nowak:1978au}, i.e.,

\be
\gamma=2.36\pm0.04, ~~~~R_n=0.189\pm0.015, ~~~~R_c=0.664\pm0.011,
\ee

The subtraction constants $a_l$ in Eq.~(\ref{eq:Our G result}) for different channels can be determined according to experimental data at the
$K^- p$ threshold, which are labeled with $Off-shell$ in Table~I. Moreover,
the corresponding subtraction constants in the original on-shell factorization approximation in Eq.~(\ref{eq:gpropdr}) are also listed in Table~I, which is labeled with $On-shell$\cite{15Oset:2002}.

\begin{table}[th] 
\begin{center}
\begin{tabular*}{\textwidth}{c@{\extracolsep{\fill}}cccccc}
\hline\hline
$ a_{l} $ & $\bar{K}N$ & $\pi\Lambda$ & $\pi\Sigma$ & $\eta\Lambda$ & $\eta\Sigma$ & $K\Xi$  \\
 \hline
On-shell & $-1.84$ & $-1.83$ & $-2.0$ & $-2.25$ & $-2.38$ & $-2.67$\\
Off-shell & $-1.1$ & $-1.6$ & $-1.9$ & $-2.7$ & $-2.6$ & $-2.8$\\
\hline\hline
\end{tabular*}
\end{center}
\caption{The subtraction constants used in the calculation with $\mu=630MeV$. The label $Off-shell$ denotes the values for the loop function in Eq.~(\ref{eq:Our G result}), where some off-shell corrections are taken into account, while the label $On-shell$ stands for those original values in the on-shell factorization approximation in Eq.~(\ref{eq:gpropdr})\cite{15Oset:2002}. }
\end{table}

The corresponding values of the energy shift $\Delta E$ and width $\Gamma$ of the 1s state of the Kaonic hydrogen, the braching ratios $R_n$ and $R_c$ defined in Eqs.~(\ref{eq:Rn}) and (\ref{eq:Rc}) reproduced with the loop function in Eq.~(\ref{eq:Our G result}) are listed in Table~II, and the
subtraction constants labeled with  $Off-shell$ are used. These values are also calculated with the $On-shell$
subtraction constants in the on-shell factorization approximation. The results show that the $Off-shell$ subtraction constants fitted with the experiment data
at the $K^- p$ threshold are reasonable.

\begin{table}[th] 
\begin{center}
\begin{tabular*}{\textwidth}{c@{\extracolsep{\fill}}cccc}
\hline\hline
 & $\Delta E\quad[eV]$ & $\Gamma \quad[eV]$ & $Rn$ & $Rc$  \\
 \hline
 Experimental data& $283\pm 36\pm 6$ & $541 \pm 89 \pm 22$ & $0.189\pm 0.015$ & $0.664\pm 0.011$ \\
On-shell & $-180.11$ & $444.14$ & $0.28$ & $0.61$ \\
Off-shell & $283.13$ & $541.06$ & $0.3$ & $0.61$ \\
\hline\hline
\end{tabular*}
\end{center}
\caption{The energy shift $\Delta E$ and width $\Gamma$ of the 1s state of the Kaonic hydrogen, the braching ratios $R_n$ and $R_c$ defined in Eqs.~(\ref{eq:Rn}) and (\ref{eq:Rc}) calculated with different
subtraction constant sets. The meanings of labels $On-shell$ and $Off-shell$ are the same as that in Table~I.}
\end{table}

Since the meson-baryon amplitude $T_{ij}$ is calculated in the isospin sectors, the subtraction constant $a_{KN}$ is supposed to take the same value both in the $K^- p$ channel and in the $\bar{K}^0 n$ channel. Thus the branching ratio $\gamma$ is always one in our calculation.


\section{The isospin $I=0$ sector}
\label{sect:I0}

We shall discuss the scattering amplitude in the isospin states, and
thus we must use the average mass for the $\pi (\pi^+,\pi^0,
\pi^-)$, $K(K^+, K^0)$, $\bar{K}(\bar{K}^0,K^-)$, $N(p,n)$,
$\Sigma(\Sigma^+, \Sigma^0, \Sigma^-)$ and $\Xi(\Xi^0, \Xi^-)$
particles.
There are four coupled states with isospin $I=0$ and strangeness
$S=-1$, which are $\bar{K}N$,$\pi\Sigma$,$\eta\Lambda$ and $K\Xi$.

The phase conventions $\vert \pi^{+}\rangle={}-\vert 1,1\rangle$,
$\vert K^{-}\rangle={}-\vert \frac{1}{2},-\frac{1}{2}\rangle$,
$\vert \Sigma^{+}\rangle={}-\vert 1,1\rangle$ and $\vert
\Xi^{-}\rangle={}-\vert \frac{1}{2},-\frac{1}{2}\rangle$ are used
for the isospin states, which are consistent with the structure of
the $\Phi$ and $B$ matrices, and then the isospin state with $I=0$
can be written as
\begin{equation} \label{eq:3.1}
\begin{aligned}
\vert \bar{K}N,I=0\rangle={}&\frac{1}{\sqrt{2}}(\bar{K}^{0}n+K^{-}p),\\
\vert \pi\Sigma,I=0\rangle={}&-\frac{1}{\sqrt{3}}(\pi^{+}\Sigma^{-}+\pi^{0}\Sigma^{0}+\pi^{-}\Sigma^{+}),\\
\vert K\Xi,I=0\rangle={}&-\frac{1}{\sqrt{2}}(K^{0}\Xi^{0}+K^{+}\Xi^{-}).\\
\end{aligned}
\end{equation}

The correspond coefficients $C_{ij}$ for the isospin states with
$I=0$ are listed in Table~\ref{TABLE2}. With these coefficients, the
amplitudes $T$ with isospin $I=0$ can be obtained by solving the
Bethe-Salpeter equation in Eq.~(\ref{eq:amplitude}).

The squared amplitude $|T|^2$ in the  $\pi \Sigma $ channel with
isospin $I=0$ on the complex $\sqrt{s}$ plane is shown in
Fig.~1.
%
Some poles are generated dynamically when the
Bethe-Salpeter equation is solved in the unitary coupled-channel
approximation.
In Fig.~1,
the poles generated dynamically
with the loop function in Eq.~(\ref{eq:Our G result}) are
labeled with $NEW$, while the label $PRE$ denotes the poles
generated with the loop function in the on-shell
factorization in Eq.~(\ref{eq:gpropdr}).
In the energy region near 1400MeV, it can be seen that there is only one pole generated dynamically in
the isospin $I=0$ sector, which locates at $1383+81i$MeV on the complex $\sqrt{s}$ plane. This pole is higher than
the $\pi \Sigma$ threshold and lies in the second Riemann sheet, and thus it can be regarded as a counterpart of
the $\Lambda(1405)$ particle in the data of Particle Data
Group(PDG)\cite{pdg}.
Actually, there is another peak generated dynamically, which is at $1435+2i$MeV on the complex $\sqrt{s}$ plane.
However, it is too low to be detected in Fig.~1. Apparently, when the off-shell correction of the loop function in the Bethe-Salpeter equation is taken into account, the peak near the $\bar{K} N$ threshold is suppressed strongly, and only one pole is detected in the region of 1400MeV. The resonance at $1383+81i$MeV
couples strongly to the $\pi \Sigma$ channel.
In the $\bar{K}N$, $\eta \Lambda$ and $K \Xi$ channels, the cases are similar to that in the $\pi \Sigma$ channel, and only one pole can be found clearly.
Furthermore, there is another pole at $1653+12i$MeV generated dynamically on the complex $\sqrt{s}$ plane,
which is lower than the $\eta \Lambda$ threshold, and lies in the third Riemann sheet.
The pole at the position of $\sqrt{S}=1653+12i$MeV can be regarded
as a candidate of the $\Lambda(1670)1/2^{-}$ resonance, which
couples strongly to the $K \Xi$ channel.

The coupling constants of them to different meson-baryon states are listed in Table.~\ref{TABLE3}, where the
label $New$ denotes the results with the loop function in
Eq.~(\ref{eq:Our G result}), while the label  $PRE$ means the
results calculated with the loop function in the on-shell
factorization, as in Eq.~(\ref{eq:gpropdr}).

\begin{table}[th]
\begin{center}
\begin{tabular*}{\textwidth}{c@{\extracolsep{\fill}}cccc}
\hline\hline
& \multicolumn{2}{c}{PRE} & \multicolumn{2}{c}{NEW} \\
\hline
$z_{R}$ & \multicolumn{2}{c}{$1390+66i$} & \multicolumn{2}{c}{$1383+81i$} \\
($I=0$) & $g_{i}$ & $|g_{i}|$ & $g_{i}$ & $|g_{i}|$ \\
\hline
$\pi\Sigma$ & $-2.5-1.5i$ & 2.9 & $-2.1-1.5i$ & 2.5 \\
$\bar{K}N$ & $1.2+1.7i$ & 2.1 & $0.8+0.8i$ & 1.1 \\
$\eta\Lambda$ & $0.0+0.8i$ & 0.8 & $-0.1+0.2i$ & 0.24  \\
$K\Xi$ & $-0.5-0.4i$ & 0.6 & $-0.4-0.5i$ & 0.6  \\

%
\hline \hline
& \multicolumn{2}{c}{PRE} & \multicolumn{2}{c}{NEW} \\
\hline
$z_{R}$ & \multicolumn{2}{c}{$1426+16i$} & \multicolumn{2}{c}{$-$}  \\
($I=0$) & $g_{i}$ & $|g_{i}|$ & $g_{i}$ & $|g_{i}|$ \\
\hline
$\pi\Sigma$ & $0.4-1.4i$ & 1.5 & $-$  & \multicolumn{1}{c}{$-$} \\
$\bar{K}N$  & $-2.5+0.9i$ & 2.7 & $-$  & \multicolumn{1}{c}{$-$}\\
$\eta\Lambda$  & $-1.4+0.2i$ & 1.4  & $-$ & \multicolumn{1}{c}{$-$} \\
$K\Xi$  & $0.1-0.3i$ & 0.4 & $-$ & \multicolumn{1}{c}{$-$} \\

%
\hline \hline
  & \multicolumn{2}{c}{PRE} & \multicolumn{2}{c}{NEW}  \\
\hline
$z_{R}$ & \multicolumn{2}{c}{$1680+20i$} & \multicolumn{2}{c}{$1653+12i$} \\
($I=0$) & $g_{i}$ & \multicolumn{1}{c}{$|g_{i}|^{2}$} & $g_{i}$ & $|g_{i}|^{2}$ \\
\hline
$\pi\Sigma$ & $-0.0-0.3i$ & \multicolumn{1}{c}{0.3} & $-0.1-0.3i$ & 0.3 \\
$\bar{K}N$ & $0.3+0.7i$ & \multicolumn{1}{c}{0.8} & $-0.0+0.7i$ & 0.7 \\
$\eta\Lambda$ & $-1.1-0.1i$ & \multicolumn{1}{c}{1.1} & $-1.1+0.2i$ & 1.1 \\
$K\Xi$ & $3.4+0.1i$ & \multicolumn{1}{c}{3.5} & $3.0-0.1i$ & 3.0 \\
\hline\hline
\end{tabular*}
\end{center}
\caption{Coupling constants to meson-baryon states in
the isospin $I=0$ sector. The label $NEW$ denotes the case
calculated from the loop function in Eq.~(\ref{eq:Our G
result}), while the label $PRE$ stands for the case of the loop
function in the on-shell factorization approximation in
Eq.~(\ref{eq:gpropdr}).}\label{TABLE3}
\end{table}

\section{The isospin $I=1$ sector}
\label{sect:I1}

 In the isospin $I=1$ sector, we have five coupled
states, $\bar{K}N$, $\pi\Sigma$, $\pi \Lambda$, $\eta \Sigma$ and
$K\Xi$.

The isospin state with $I=1$ can be written as
\begin{equation} \label{eq:I1}
\begin{aligned}
%
\vert \bar{K}N,I=1\rangle={}&\frac{1}{\sqrt{2}}(\bar{K}^{0}n-K^{-}p),\\
\vert \pi\Sigma,I=1\rangle={}&\frac{1}{\sqrt{2}}(\pi^{-}\Sigma^{+}-\pi^{+}\Sigma^{-}),\\
\vert
K\Xi,I=1\rangle={}&\frac{1}{\sqrt{2}}(K^{0}\Xi^{0}-K^{+}\Xi^{-}).
\end{aligned}
\end{equation}

The coefficients $C_{ij}$ for the isospin states with $I=1$ can be
constructed by using Eq.~(\ref{eq:I1}) and Table~\ref{TABLE1}, which
are given in Table~\ref{TABLE2-2}.

There is only one peak of squared amplitudes $|T|^2$ detected at
$1570+i244$MeV on the complex $\sqrt{s}$ plane, which is lower than $\eta \Sigma$ threshold and lies in the fourth
Riemann sheet.
The resonance at $1570+i244$MeV is similar to the $\Sigma(1580)3/2-$
state in the $PDG$ data. Nevertheless, in the S-wave approximation,
the total angular momentum of the resonance at $1570+i244$MeV is
$J=1/2$, and its parity is negative, furthermore, the decay width is
too large, so it can not be $\Sigma(1580)3/2-$ apparently.
The resonance at $1570+i244$MeV is more possible to correspond to
the $\Sigma(1620)1/2-$ state in the $PDG$ data although its mass is
lower about $50$MeV than the latter.
From Table~\ref{TABLE4}, we can know that this resonance couples to
the $K\Xi$ channel strongly.

The squared amplitude $|T|^2$ in the $\pi \Sigma$ channel with
strangeness $S=-1$ and isospin $I=1$ as a function of the total
energy $\sqrt{s}$ in the center of mass frame is depicted in
Fig.~2, where the label $PRE$ denotes the results
calculated with the loop function in the on-shell factorization, and
the label $NEW$ means those of the loop function in Eq.~(\ref{eq:Our G result}).
The pole positions are different in these two schemes.
%
%

\begin{table}[th]
\begin{center}
\begin{tabular*}{\textwidth}{c@{\extracolsep{\fill}}cccc}
\hline\hline
  & \multicolumn{2}{c}{PRE} & \multicolumn{2}{c}{NEW}\\
\hline
$z_{R}$ & \multicolumn{2}{c}{$1579+264i$} & \multicolumn{2}{c}{$1570+244i$} \\
($I=1$) & $g_{i}$ & $|g_{i}|^{2}$ &$g_{i}$ & $|g_{i}|^{2}$ \\
\hline
$\pi\Lambda$ & $1.4+1.5i$ & 2.0 & $1.3+1.4i$ & 1.9 \\
$\pi\Sigma$ & $-2.2-1.5i$ & 2.7 & $-2.2-1.2i$ & 2.5 \\
$\bar{K}N$ & $-1.1-1.1i$ & 1.6 & $-1.0-1.2i$ & 1.6 \\
$\eta\Sigma$ & $1.2+1.4i$ & 1.9 & $1.1+1.3i$ & 1.7 \\
$K\Xi$ & $-2.5-2.4i$ & 3.5 & $-2.4-2.1i$ & 3.2 \\
\hline\hline
\end{tabular*}
\end{center}
\caption{Same as Table~\ref{TABLE3} but for the isospin $I=1$
sector. }\label{TABLE4}
\end{table}

\section{Discussion and conclusion}
\label{sect:summary}

In this article, the formula of the loop function in the
Bethe-Salpeter equation is deduced in the dimensional
regularization scheme. Comparing with the loop function in the
on-shell factorization approximation, the relativistic kinetic effect and
off-shell corrections are taken into account in the
revised scheme.

The interaction between the pseudoscalar meson and the baryon
is studied in the strangeness $S=-1$ sector. According to the experimental
data at the $K^- p$ threshold, the subtraction constants of the loop function in the Bethe-Salpeter equation are determined, and
some resonances are
generated dynamically in the unitary coupled-channel approximation.
When the off-shell correction of the loop function is taken into account, the squared amplitude is suppressed strongly, and only one resonance peak is detected in the 1400MeV region, which is above the $\pi \Sigma$ threshold and is assumed to be associated to the $\Lambda(1405)$ particle.
Moreover, the coupling constants of the resonance to meson-baryon states are similar to those in
Ref.~\cite{8Jido:2003} correspondingly, and the resonance in the 1400MeV region couples strongly to the $\pi \Sigma$ channel.

\section*{Acknowledgments}
We would like to thank Han-Qing Zheng, Eulogio Oset and En Wang for
useful discussions.


\section*{Appendix 1: A loop function in the dimensional
 regularization scheme}

In the dimensional regularization scheme, the loop function in the
Bethe-Salpeter equation can be deduced according to the Passarino-Veltman procedure\cite{Passarino},


\be
G_l(s)=i\int \frac{d^4 q}{(2\pi)^4} \frac{\rlap{/} q +M_l}{q^2-M_l^2+i\epsilon } \frac{1}{(P-q)^2-m_l^2+i\epsilon}
=-\rlap{/} P A(s)+\frac{1}{2} G^\prime(s), \nn
~~~~\hfill(A1)
\ee

where
 \ba
G^\pr(s)&=&i \int \frac{d^4 q}{(2\pi)^4} \frac{2M_l}{q^2-M_l^2+i\epsilon } \frac{1}{(P-q)^2-m_l^2+i\epsilon} \nn \\
&=& \frac{2M_l}{16 \pi^2} \left\{ a_l + \ln \frac{M_l^2}{\mu^2}
+ \frac{m_l^2-M_l^2 + s}{2s} \ln \frac{m_l^2}{M_l^2} + \right.
\nonumber \\ & &  \phantom{\frac{2 M}{16 \pi^2}} +
\frac{\bar{q}_l}{\sqrt{s}} \left[
\ln(s-(M_l^2-m_l^2)+2\bar{q}_l\sqrt{s})+
\ln(s+(M_l^2-m_l^2)+2\bar{q}_l\sqrt{s}) \right. \nonumber  \\
& & \left. \phantom{\frac{2 M}{16 \pi^2} +
\frac{\bar{q}_l}{\sqrt{s}}} \left. \hspace*{-0.3cm}-
\ln(-s+(M_l^2-m_l^2)+2\bar{q}_l\sqrt{s})-
\ln(-s-(M_l^2-m_l^2)+2\bar{q}_l\sqrt{s}) \right] \right\} \ , \nn \hfill(A2) \ea
and
 \ba
P^\mu A(s) &=& -i  \int \frac{d^4 q}{(2\pi)^4} \frac{
q^\mu}{q^2-M_l^2+i\epsilon } \frac{1}{(P-q)^2-m_l^2+i\epsilon}. \nn
\ea

Since $-2 P \cdot q~=~[(P-q)^2-m_l^2]-[q^2-M_l^2]+m_l^2-M_l^2-P^2$,
\ba
-2P^2 A(s) &=& -i  \int \frac{d^4 q}{(2\pi)^4} \frac{ -2 P \cdot
q}{q^2-M_l^2+i\epsilon } \frac{1}{(P-q)^2-m_l^2+i\epsilon} \nn
\\
&=&-i  \int \frac{d^4 q}{(2\pi)^4} \frac{
[(P-q)^2-m_l^2]-[q^2-M_l^2]+m_l^2-M_l^2-P^2}{[q^2-M_l^2+i\epsilon ]
[(P-q)^2-m_l^2+i\epsilon]} \nn \\
&=&-i  \int \frac{d^4 q}{(2\pi)^4} \frac{ 1}{q^2-M_l^2+i\epsilon}
-i  \int \frac{d^4 q}{(2\pi)^4} \frac{ -1}{ (P-q)^2-m_l^2+i\epsilon}
\nn \\
&&  -i  \int \frac{d^4 q}{(2\pi)^4} \frac{
m_l^2-M_l^2-P^2}{[q^2-M_l^2+i\epsilon ] [(P-q)^2-m_l^2+i\epsilon]} \nn
 \\
&=&
-\frac{M_l^2}{16\pi^2} \left(1+a_l+\ln(\frac{M_l^2}{\mu}) \right)
+\frac{m_l^2}{16\pi^2} \left(1+a_l+\ln(\frac{m_l^2}{\mu}) \right)
-\frac{(P^2+M_l^2-m_l^2)}{2M_l}(-G^\prime(s)).  \nn \ea
Thus the formula of $A(s)$ can be written as
 \be
A(s)=\frac{1}{-2P^2}
\{-\frac{M_l^2}{16\pi^2} \left(1+a_l+\ln(\frac{M_l^2}{\mu}) \right)
+\frac{m_l^2}{16\pi^2} \left(1+a_l+\ln(\frac{m_l^2}{\mu}) \right)
-\frac{(P^2+M_l^2-m_l^2)}{2M_l}(-G^\prime(s)) \},\nn \hfill(A3)\ee with $P^2=s$.

By replacing the $G^\pr(s)$ and $A(s)$ in Eq.~(A1) with
 Eqs.~(A2) and (A3),
respectively, the formula of the loop function in Eq.~(\ref{eq:Our
G}) is obtained.

\newpage

\section*{Appendix 2: The coefficient $C_{ij}$ in the pseudoscalar meson-baryon
octet interaction in the $S=-1$ channel}

\begin{table}[hbt]
\begin{center}
\begin{tabular}{ c c c c c c c c c c c }
\hline
 & $K^{-}p$ & $\bar{K}^{0}n$ & $\pi^{0}\Lambda$ & $\pi^{0}\Sigma^{0}$ & $\eta\Lambda$ & $\eta\Sigma^{0}$ & $\pi^{+}\Sigma^{-}$ & $\pi^{-}\Sigma^{+}$ & $K^{+}\Xi^{-}$ & $K^{0}\Xi^{0}$ \\
\hline
$K^{-}p$ & 2 & 1 & $\frac{\sqrt{3}}{2}$ & $\frac{1}{2}$ & $\frac{3}{2}$ & $\frac{\sqrt{3}}{2}$ & 0 & 1 & 0 & 0 \\
$\bar{K}^{0}n$ &  & 2 & $-\frac{\sqrt{3}}{2}$ & $\frac{1}{2}$ & $\frac{3}{2}$ & $-\frac{\sqrt{3}}{2}$ & 1 & 0 & 0 & 0 \\
$\pi^{0}\Lambda$ &  &  & 0 & 0 & 0 & 0 & 0 & 0 & $\frac{\sqrt{3}}{2}$ & $-\frac{\sqrt{3}}{2}$ \\
$\pi^{0}\Sigma^{0}$ &  &  &  & 0 & 0 & 0 & 2 & 2 & $\frac{1}{2}$ & $\frac{1}{2}$ \\
$\eta\Lambda$ &  &  &  &  & 0 & 0 & 0 & 0 & $\frac{3}{2}$ & $\frac{3}{2}$ \\
$\eta\Sigma^{0}$ &  &  &  &  &  & 0 & 0 & 0 & $\frac{\sqrt{3}}{2}$ & $-\frac{\sqrt{3}}{2}$ \\
$\pi^{+}\Sigma^{-}$ &  &  &  &  &  &  & 2 & 0 & 1 & 0 \\
$\pi^{-}\Sigma^{+}$ &  &  &  &  &  &  &  & 2 & 0 & 1 \\
$K^{+}\Xi^{-}$ &  &  &  &  &  &  &  &  & 2 & 1 \\
$K^{0}\Xi^{0}$ &  &  &  &  &  &  &  &  &  & 2 \\
\hline
\end{tabular}
\end{center}
\caption{  The coefficient $C_{ij}$ in the pseudoscalar meson-baryon
octet interaction in the $S=-1$ channel, $C_{ij}=C_{ji}$. }
  \label{TABLE1}
\end{table}

\begin{table}[hbt]
\begin{center}
\begin{tabular}{*{5}{p{.15\textwidth}}}
\hline
 & $\bar{K}N$ & $\pi\Sigma$ & $\eta\Lambda$ & $K\Xi$ \\
\hline
$\bar{K}N$ & 3 & $-\sqrt{\frac{3}{2}}$ & $\frac{3}{\sqrt{2}}$ & 0 \\
$\pi\Sigma$ &  & 4 & 0 & $\sqrt{\frac{3}{2}}$ \\
$\eta\Lambda$ &  &  & 0 & $-\frac{3}{\sqrt{2}}$ \\
$K\Xi$ &  &  &  & 3\\
\hline
\end{tabular}
\end{center}
\caption { The coefficients $C_{ij}$ for the isospin states with
$I=0$, $C_{ij}=C_{ji}$. }
 \label{TABLE2}
\end{table}

\begin{table}[hbt]
\begin{center}
\begin{tabular}{*{6}{p{.10\textwidth}}}
\hline
 & $\bar{K}N$ & $\pi\Sigma$ & $\pi\Lambda$ & $\eta \Sigma$ & $K\Xi$ \\
\hline
$\bar{K}N$    & $1$ & $-1$ & $-\sqrt{\frac{3}{2}}$ & $-\sqrt{\frac{3}{2}}$  & $0$\\
$\pi\Sigma$   &  & $2$ & $0$ & $0$  & $1$\\
$\pi\Lambda$  &  &  & $0$ & $0$  & $-\sqrt{\frac{3}{2}}$\\
$\eta \Sigma$ &  &  &  & $0$  & $-\sqrt{\frac{3}{2}}$\\
$K\Xi$        &  &  &  &   & $1$\\
\hline
\end{tabular}
\end{center}
\caption { The coefficients $C_{ij}$ for the isospin states with
$I=1$, $C_{ij}=C_{ji}$. }
 \label{TABLE2-2}
\end{table}

\newpage

\leftline{\Large {\bf Figure Captions}}
\parindent = 2 true cm
\parskip 1 cm
\begin{description}

\item[Fig. 1] \label{FIG1} Comparison of poles in the strangeness $S=-1$ and isospin
$I=0$ sector. $NEW$ denotes the case calculated from the
loop function in Eq.~(\ref{eq:Our G result}), while $PRE$ stands for
the case of the loop function in the on-shell factorization
approximation in Eq.~(\ref{eq:gpropdr}).\\


\par

\item[Fig. 2] \label{FIG3} Same as Fig.~1 but for the strangeness $S=-1$ and
isospin $I=1$ sector.
\\

%

\end{description}

\end{document}